%% file: johno.tex
\documentclass[review,onefignum,onetabnum]{siamart171218}

\input{johno_shared}

\ifpdf
\hypersetup{
  pdftitle={Prediction-based compensation for gate on/off latency},
  pdfauthor={H. Johno, M. Saito, and H. Onishi}
}
\fi

\begin{document}

\maketitle

% REQUIRED
\begin{abstract}
During respiratory-gated radiotherapy (RGRT), gate on and off latencies cause deviations of gating windows, possibly leading to delivery of low- and high-dose radiations to tumors and normal tissues, respectively.
Currently, there are no RGRT systems that have definite tools to compensate for the delays.
To address the problem, we propose a framework consisting of two steps: 1) multi-step-ahead prediction and 2) prediction-based gating.
For each step, we have devised a specific algorithm to accomplish the task.
Numerical experiments were performed using respiratory signals of a phantom and ten volunteers, and our prediction-based RGRT system exhibited superior performance in more than a few signal samples.
In some, however, signal prediction and prediction-based gating did not work well, maybe due to signal irregularity and/or baseline drift.
The proposed approach has potential applicability in RGRT, and further studies are needed to verify and refine the constituent algorithms.
\end{abstract}

% REQUIRED
\begin{keywords}
respiratory-gated radiotherapy, gate on/off latency, gating window, multi-step-ahead prediction, prediction-based gating
\end{keywords}

% REQUIRED３
%\begin{AMS}
%00A06, 00A69
%\end{AMS}

\section{Introduction}
\begin{figure}[htbp]
\centering
\subfloat[Ideal gating windows]{\label{fig:idealRGRT}\includegraphics[width=0.35\textwidth]{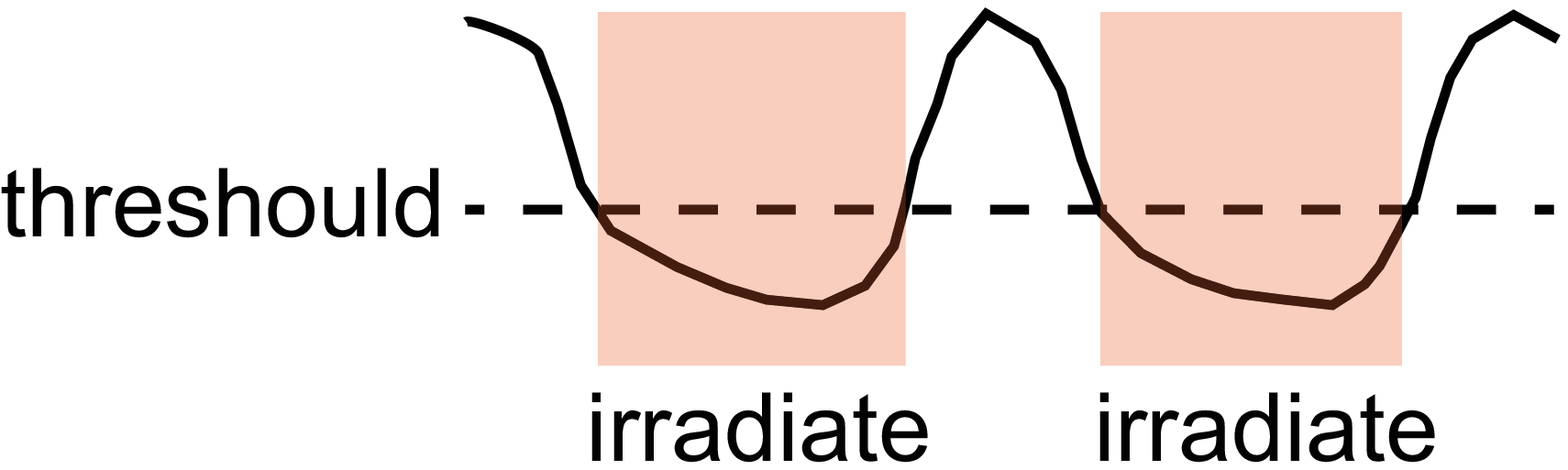}}
\hspace{0.25 in}
\subfloat[Shifted gating windows]{\label{fig:convRGRT}\includegraphics[width=0.35\textwidth]{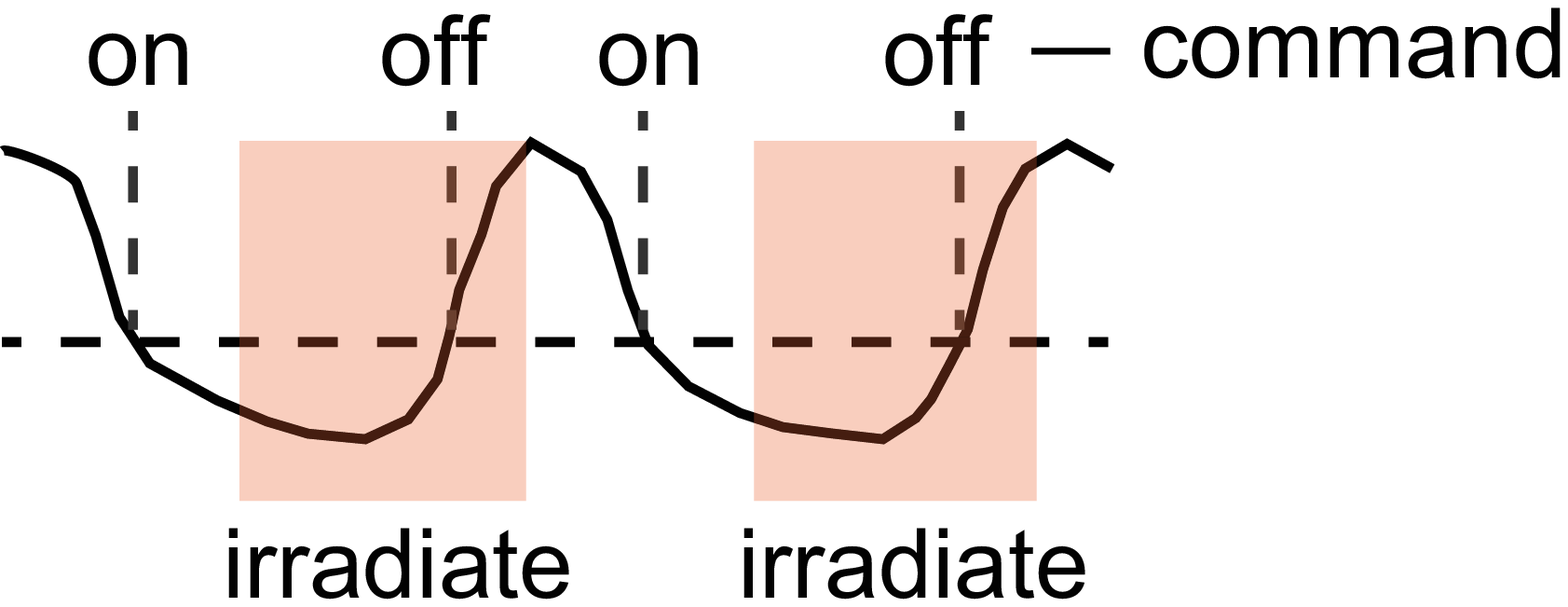}}
\caption{Problem with conventional RGRT. (a) Given a respiratory signal and a gating threshold, gating windows should ideally be the time when the signal is lower than the threshold. (b) In conventional RGRT, gate on and off commands are sent just when the signal is lower and higher than the threshold, respectively. Hence gate on and off delays cause deviations of gating windows.}
\label{fig:idconv_RGRT}
\end{figure}

Respiratory-gated radiotherapy (RGRT) is a widely employed means of treating tumors that move with respiration \cite{kubo96, vedam01, keall06}. 
In RGRT, radiation is administered within particular phases of patient's breathing cycle (called as gating windows), which are determined by monitoring respiratory motion in the form of a respiratory signal using either external or internal markers.
Note that, although there are some options for RGRT (e.g., whether to choose amplitude-based or phase-based gating and whether to gate during inhalation or exhalation), this study focuses only on amplitude-based gating during exhalation, which is a common setting in clinical practice.
Several RGRT systems have been developed, and some take considerable time from the detection of a signal change to the execution of a gate on/off command (\cref{tab:delay}).
\begin{table}[tbhp]
{\footnotesize
\caption{Gate on and off latencies of some gating systems}\label{tab:delay}
\begin{center}
\begin{tabular}{|l|l|r|r|r|} \hline
Monitor & Linac & Gate on delay & Gate off delay & Reference\\ \hline
Abches (APEX) & Elekta Synergy & 336 ms & 88 ms & \cite{saito18} \\
AlignRT (VisionRT) & Varian Clinac iX & 356 ms & 529 ms & \cite{wiersma16} \\
Calypso (Varian) & Varian Clinac iX & 209 ms & 60 ms & \cite{wiersma16} \\
Catalyst (C-RAD) & Elekta Synergy & 851 ms & 215 ms & \cite{freislederer15} \\
\hline
\end{tabular}
\end{center}
}
\end{table}
The gate on/off latency causes deviations of gating windows in conventional RGRT (\cref{fig:idconv_RGRT}), possibly leading to delivery of low- and high-dose radiation to tumor and normal tissues, respectively.
At present, there are no RGRT systems that have definite techniques to compensate for the delays.
Therefore, here, we propose a prediction-based system to address the problem.

This paper is organized as follows. 
The devised framework is described in \cref{sec:methods}, experimental results are in \cref{sec:exp}, and the conclusions follow in \cref{sec:ccl}.

\section{Methods}\label{sec:methods}
In this section, we describe our new approach to compensate for gate on/off latency.
This consists of two steps: 1) multi-step-ahead prediction and 2) prediction-based gating.
\subsection{Multi-step-ahead prediction}\label{sec:forecast}
Several prediction algorithms for respiratory signals have been proposed, and most of them adopt single-output strategies \cite{lee14, taieb10}.
However, in our framework, multiple-output multi-step-ahead prediction is required.
Therefore we have devised an algorithm for this purpose.

A respiratory signal is regarded as a sequence 
\begin{equation*}
\{x_t\},\ t=0,1,2,\ldots,
\end{equation*}
of equally spaced time-series observations in a space $\mathcal{X}$, with a time interval of $\varDelta\tau$ seconds (s), where $\varDelta\tau>0$.
Let $n$ and $m$ be positive integers.
For each time point $t\ge n$, multi-step-ahead prediction aims to forecast the $m$-tuple $(x_t,\ldots,x_{t+m-1})$ of subsequent observations, given the previous $n$-tuple $(x_{t-n},\ldots,x_{t-1})$.
Hence our goal here is to form a predictor mapping on $\mathcal{X}^n$ to $\mathcal{X}^m$.
Suppose $\mathcal{X}^n$ is a metric space with a metric $d_n$.
Let us have a learning set $\mathcal{L} = \{(\tilde{\mathbf{x}}_i, \tilde{\mathbf{y}}_i)\in \mathcal{X}^n\times\mathcal{X}^m\}$, where $i$ ranges over some finite totally ordered set (see \cref{sec:learning_set} for an example of the learning set preparation).
Then for a test tuple $\mathbf{x}\in\mathcal{X}^n$, we predict the next $m$-tuple as 
\begin{equation*}
\Psi_\mathcal{L}(\mathbf{x}) = \tilde{\mathbf{y}}_p,
\end{equation*}
where $p$ is the largest index such that $d_n(\mathbf{x}, \tilde{\mathbf{x}}_p) \le d_n(\mathbf{x}, \tilde{\mathbf{x}}_i)$ for all $i$.
Throughout this paper, we suppose that $\mathcal{X} = \mathbb{R}$ and $\mathcal{X}^k$, which equals $\mathbb{R}^k$ ($k=1,2,3,\ldots$), is a real $k$-space with the Euclidean metric, i.e., 
\begin{equation*}
d_k(\mathbf{a}, \mathbf{b}) = \sqrt{\sum_{j=0}^{k-1}(a_j-b_j)^2}
\quad \left(
\begin{aligned}
&\mathbf{a}=(a_0,\ldots,a_{k-1})\in\mathbb{R}^k\\
&\mathbf{b}=(b_0,\ldots,b_{k-1})\in\mathbb{R}^k
\end{aligned}\right).
\end{equation*}

\subsection{Prediction-based RGRT}\label{sec:pred_gate}
Let $x_t\in\mathbb{R}^1$ ($t\ge n$) be the current observation, $\beta\in\mathbb{R}^1$ a gating threshold, and $m_1$ and $m_0$ the numbers of time points corresponding to gate on and off delays, respectively.
Given learning sets $\mathcal{L}_1\subset\mathbb{R}^n\times\mathbb{R}^{2m_1+1}$ and $\mathcal{L}_0\subset\mathbb{R}^n\times\mathbb{R}^{2m_0+1}$ (see \cref{sec:learning_set} for an example of the learning set construction), the function $G_{\mathcal{L}_1,\mathcal{L}_0}$ defined below is used for a prediction-based gating.
\begin{enumerate}
\item Case $m_1\ge m_0$:
\begin{align*}
G_{\mathcal{L}_1,\mathcal{L}_0}(t,\beta) =
\begin{cases}
\ 1 & \mbox{if}\ \,\xi_{2m_1+1,\beta}\left(\Psi_{\mathcal{L}_1}(\mathbf{x}_{t})\right)<0\ \,\mbox{or}\ \,\xi_{2m_0+1,\beta}\left(\Psi_{\mathcal{L}_0}(\mathbf{x}_{t})\right)<0 \\
\ 0 & \mbox{otherwise}
\end{cases},
\end{align*}
\item Case $m_1< m_0$:
\begin{align*}
G_{\mathcal{L}_1,\mathcal{L}_0}(t,\beta) =
\begin{cases}
\ 1 & \mbox{if}\ \,\xi_{2m_1+1,\beta}\left(\Psi_{\mathcal{L}_1}(\mathbf{x}_{t})\right)<0\ \,\mbox{and}\ \,\xi_{2m_0+1,\beta}\left(\Psi_{\mathcal{L}_0}(\mathbf{x}_{t})\right)<0 \\
\ 0 & \mbox{otherwise}
\end{cases},
\end{align*}
\end{enumerate}
where $\xi_{m,\beta}: \mathbb{R}^m\to\mathbb{Z}$ (the set of integers) is defined by
\begin{equation*}
\xi_{m,\beta}\left((a_0,\ldots,a_{m-1})\right) = \sum_{k=0}^{m-1}\mathrm{sgn} \left(a_k-\beta \right)
\end{equation*}
and $\mathbf{x}_{t} = (x_{t-n},\ldots,x_{t-1})\in\mathbb{R}^n$.
Note that $\mathrm{sgn}: \mathbb{R}^1\to\{-1,0,1\}$ denotes the signum function, i.e.,
\begin{equation*}
\mathrm{sgn}(a) = 
\begin{cases}
\ 1 & \mbox{if}\ \,a>0\\
\ 0 & \mbox{if}\ \,a=0\\
\ -1 & \mbox{if}\ \,a<0
\end{cases}.
\end{equation*}

In our prediction-based RGRT system (pRGRT), gate on command is sent if $G_{\mathcal{L}_1,\mathcal{L}_0}(t,\beta) = 1$, while gate off command is sent if $G_{\mathcal{L}_1,\mathcal{L}_0}(t,\beta) = 0$.

\subsection{Construction of a learning set}\label{sec:learning_set}
To begin with, a respiratory signal tuple $(x_0,\ldots,x_{N-1})\in \mathbb{R}^N$ is smoothed using the finite Fourier transform \cite{nicholson71}.
In detail, the mapping $\Phi_{\alpha,N}: \mathbb{R}^N\to\mathbb{R}^N$ $(\alpha\in\mathbb{R}^1)$ defined below is applied for the smoothing.
\begin{subequations}
\begin{align}
&\Phi_{\alpha,N}\left((x_0,\ldots,x_{N-1})\right) = \left(\tilde{x}_0,\ldots,\tilde{x}_{N-1}\right),\label{eq:Phi1}\\
&\left(s_0,\ldots,s_{N-1}\right) = \mathcal{F}_N\left(W((x_0,\ldots,x_{N-1}))\right),\label{eq:Phi4} \\
&\left(u_0,\ldots,u_{N-1}\right) = \mathcal{F}_N^{-1}\left(F_\alpha((s_0,\ldots,s_{N-1}))\right),\label{eq:Phi3} \\
&\left(\tilde{x}_0,\ldots,\tilde{x}_{N-1}\right) = W^{-1}\left(R\left((u_0,\ldots,u_{N-1})\right)\right),\label{eq:Phi2}
\end{align}
\end{subequations}
where $\mathcal{F}_N$ is the finite Fourier transform on $\mathbb{C}^N$ (a complex $N$-space) defined by
\begin{align*}\label{FT}
&\mathcal{F}_N\left((a_0,\ldots,a_{N-1})\right) = \left(\hat{a}_0,\ldots,\hat{a}_{N-1}\right), \\
&\hat{a}_k = \sum_{j=0}^{N-1}a_j\exp\left(-\frac{2\pi\sqrt{-1}jk}{N}\right)
\quad (k=0,\ldots,N-1),
\end{align*}
while its inverse is given by 
\begin{align*}\label{IFT}
&\mathcal{F}_N^{-1}\left((a_0,\ldots,a_{N-1})\right) = \left(\check{a}_0,\ldots,\check{a}_{N-1}\right), \\
&\check{a}_k = \frac{1}{N}\sum_{j=0}^{N-1}a_j\exp\left(\frac{2\pi\sqrt{-1}jk}{N}\right)
\quad (k=0,\ldots,N-1),
\end{align*}
$W: \mathbb{R}^N\to\mathbb{R}^N$ is defined by 
\begin{align*}
&W\left((x_0,\ldots,x_{N-1})\right) = \left(w_0x_0,\ldots,w_{N-1}x_{N-1}\right), \\
&w_k = 0.54 - 0.46\cos\left(\frac{2\pi k}{N-1}\right)\quad (k=0,\ldots,N-1),
\end{align*}
while its inverse is given by
\begin{equation*}
W^{-1}\left((x_0,\ldots,x_{N-1})\right) = \left(x_0/w_0,\ldots,x_{N-1}/w_{N-1}\right), 
 \end{equation*}
$F_\alpha: \mathbb{C}^N\to\mathbb{C}^N$ is defined by 
\begin{align*}
&F_\alpha((s_0,\ldots,s_{N-1})) = \left(\tilde{s}_0,\ldots,\tilde{s}_{N-1}\right), \\
&\tilde{s}_k = 
\begin{cases}
\ 0 & \mbox{if}\ \left|k-N/2\right| < N/2-\alpha \\
\ s_k & \mbox{otherwise}\quad
\end{cases}\quad (k=0,\ldots,N-1),
\end{align*}
and $R:\mathbb{C}^N\to\mathbb{R}^N$ is by
\begin{equation*}
R\left((u_0,\ldots,u_{N-1})\right) = \left(\mathrm{Re}(u_0),\ldots,\mathrm{Re}(u_{N-1})\right).
\end{equation*}
Note that $W$ defined above is called the Hamming window \cite{stoica05}.
The parameter $\alpha\in\mathbb{R}^1$ can be set freely, e.g., we set 
\begin{equation*}
\alpha=N\varDelta\tau f\quad \left(0\le f\le \frac{1}{2\varDelta\tau}\right)
\end{equation*}
 to filter out signal components with frequencies larger than $f$ hertz (Hz).

For a signal tuple $(x_0,\ldots,x_{N-1})\in \mathbb{R}^N$, 
\begin{equation*}
\left(\tilde{x}_0,\ldots,\tilde{x}_{N-1}\right)= \Phi_{\alpha,N}\left((x_0,\ldots,x_{N-1})\right)
\end{equation*}
is called the \textit{smoothed signal tuple} and used to construct a learning set $\{(\tilde{\mathbf{x}}_i, \tilde{\mathbf{y}}_i)\}\subset \mathbb{R}^n\times\mathbb{R}^m$ ($n+m\le N$) by putting
\begin{align*}
&\tilde{\mathbf{x}}_i = \left(\tilde{x}_i,\ldots,\tilde{x}_{i+n-1}\right),\\
&\tilde{\mathbf{y}}_i = \left(\tilde{x}_{i+n},\ldots,\tilde{x}_{i+n+m-1}\right)
\end{align*}
for $i=0,\ldots,N-n-m$.

\section{Numerical results and discussion}\label{sec:exp}
To validate the devised algorithms, respiratory signals of a dynamic thoracic phantom (CIRS, Virginia, USA) and ten healthy volunteers were measured with Abches (APEX Medical, Inc., Tokyo, Japan), which is a respiration-monitoring device developed by Onishi et al. \cite{onishi10} and routinely used in our university hospital.
Note that, for simplicity, we supposed that $\varDelta \tau=0.03$, although the actual time intervals were not precisely equal to 0.03 s.
Signal values were given in the unit of mm.

\subsection{Smoothing of a respiratory signal}\label{sec:model}
\begin{figure}[htbp]
\centering
\includegraphics[width=0.95\textwidth]{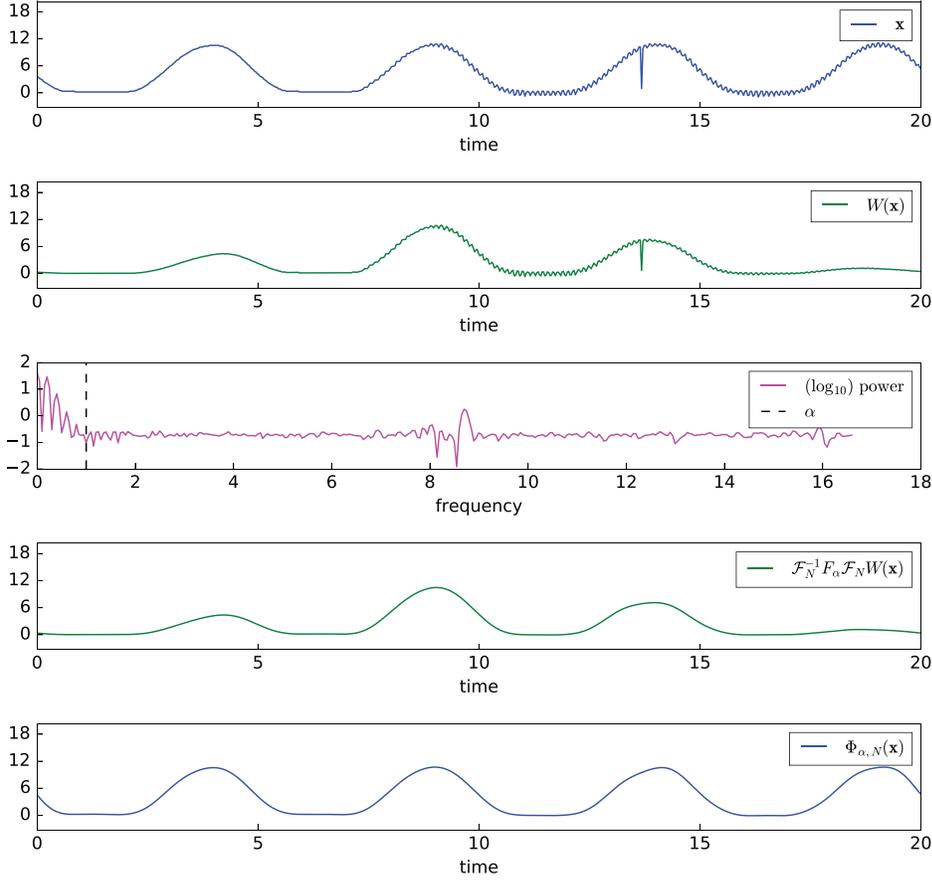}
\caption{Smoothing of a respiratory signal. The phantom's signal was measured for 20 \textup{s} and an artificial noise was added (13.65-13.7 \textup{s}), forming a signal tuple $\mathbf{x} = (x_0,\ldots,x_{N-1})$. Then $\Phi_{\alpha, N}(\mathbf{x})$ was calculated (see equations \cref{eq:Phi1,eq:Phi2,eq:Phi3,eq:Phi4}), setting $\alpha=N\varDelta\tau$ to filter out high frequency ($>1$ \textup{Hz}) components. 
Note that $\mathrm{power}$ indicates $(|s_0|,\ldots,|s_{\lfloor(N-1)/2\rfloor}|)$, where $(s_0,\ldots,s_{N-1}) = \mathcal{F}_N(W((x_0,\ldots,x_{N-1})))$, $|s|$ denotes the absolute value of $s\in\mathbb{C}^1$, and $\lfloor a \rfloor$ is the largest integer smaller than or equal to $a\in\mathbb{R}^1$.
The units of signal value, time, and frequency are \textup{mm}, \textup{s}, and \textup{Hz}, respectively.}
\label{fig:modeling}
\end{figure}

To test the algorithm of smoothing a respiratory signal, the phantom's signal was measured for 20 s (667 time points) and an artificial noise was added (13.65--13.7 \textup{s}), forming a signal tuple $\mathbf{x} = (x_0,\ldots,x_{N-1})$.
Then $\Phi_{\alpha, N}(\mathbf{x})$ was calculated (see equations \cref{eq:Phi1,eq:Phi2,eq:Phi3,eq:Phi4}), setting $\alpha=N\varDelta\tau$ to filter out high frequency ($>1$ Hz) components.
As shown in \cref{fig:modeling}, we succeeded in removing noisy components of $\mathbf{x}$.

\subsection{Prediction of a respiratory signal}\label{sec:prediction}
\begin{figure}[htbp]
\centering
\includegraphics[width=0.95\textwidth]{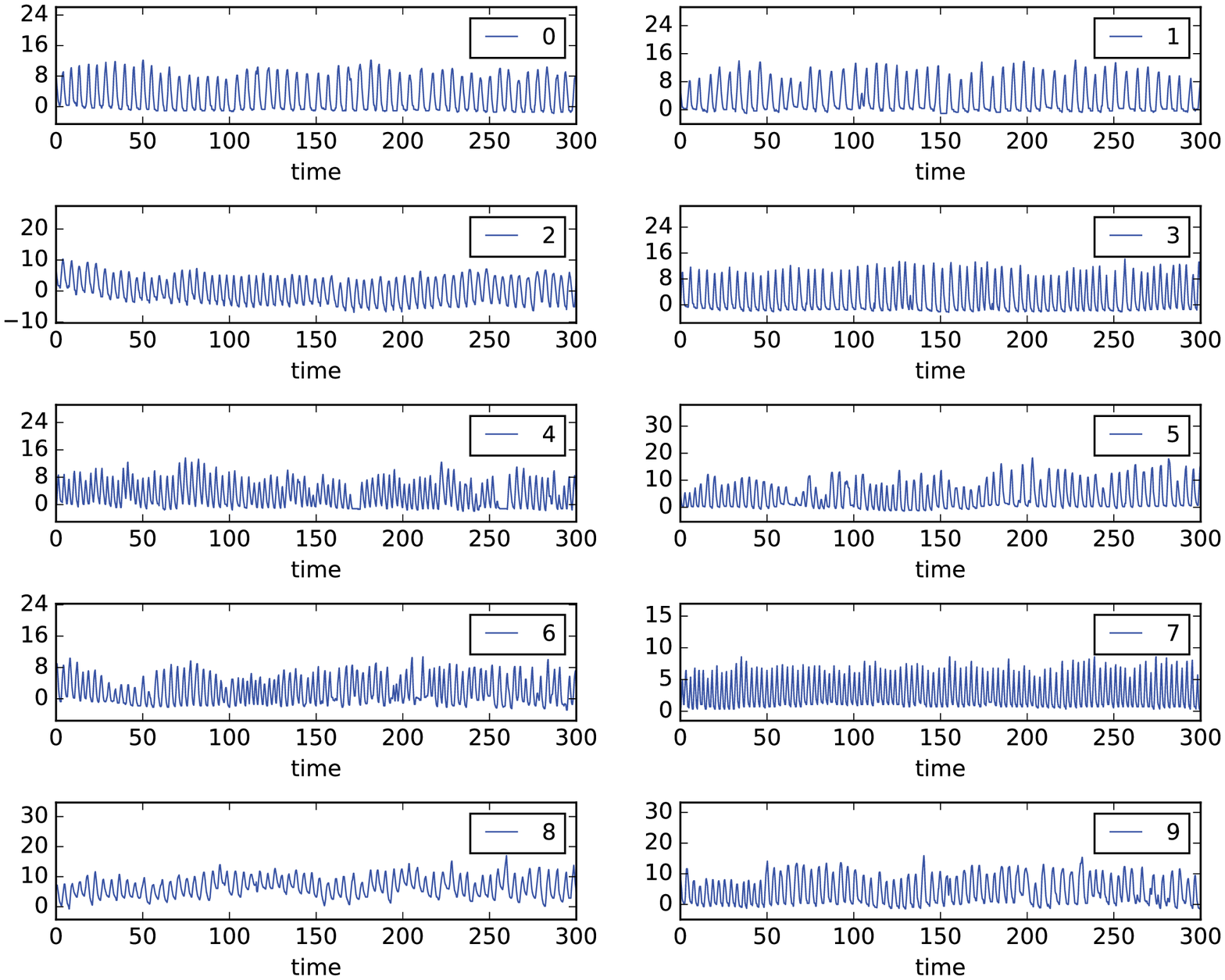}
\caption{Respiratory signal samples of ten volunteers measured for 300 \textup{s}. The units of signal value and time are \textup{mm} and \textup{s}, respectively.}
\label{fig:raw}
\end{figure}

\begin{figure}[htbp]
\centering
\includegraphics[width=0.53\textwidth]{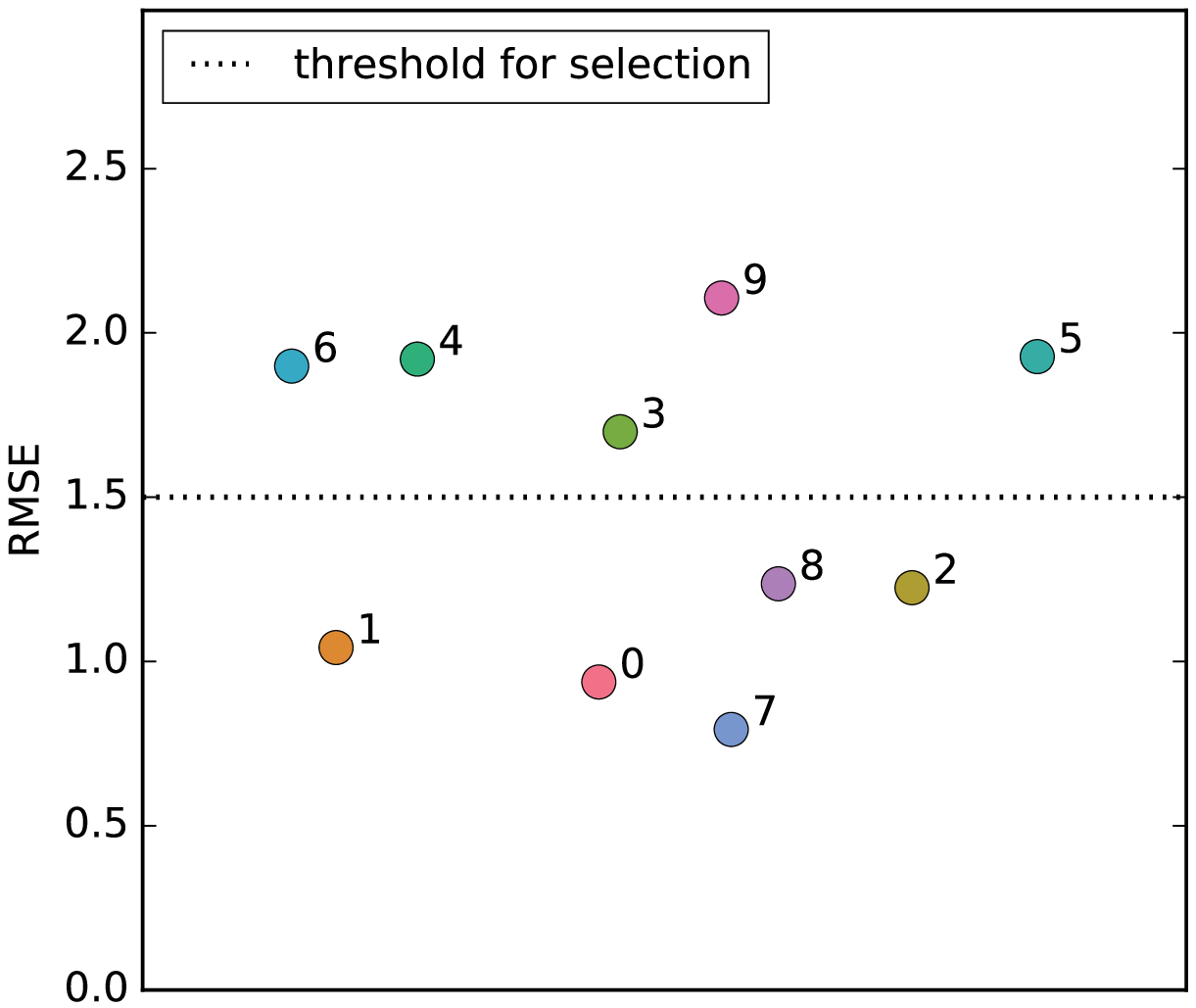}
\caption{Prediction errors for the ten samples. For each sample, RMSE (\textup{mm}) value was calculated (see \cref{sec:prediction}) and is plotted here.}
\label{fig:rmse}
\end{figure}

\begin{figure}[htbp]
\centering
\includegraphics[width=0.64\textwidth]{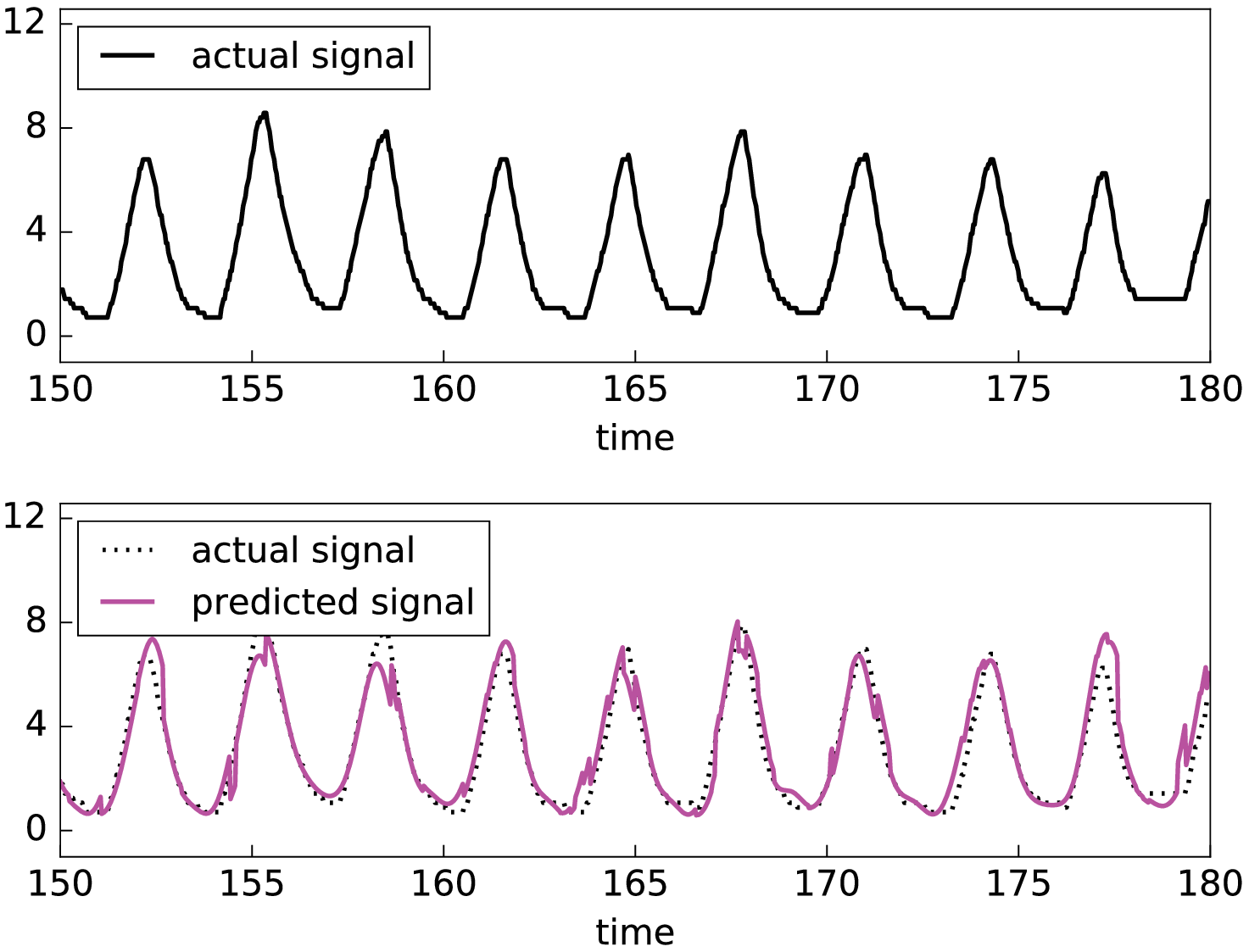}
\caption{Representative prediction result. Predicted signal corresponds to $\hat{x}_{5000},\ldots,\hat{x}_{5999}$ of the serial prediction trials (see \cref{sec:prediction}) using the sample numbered 7. The units of signal value and time are \textup{mm} and \textup{s}, respectively.}
\label{fig:predict}
\end{figure}

\begin{figure}[htbp]
\centering
\includegraphics[width=0.64\textwidth]{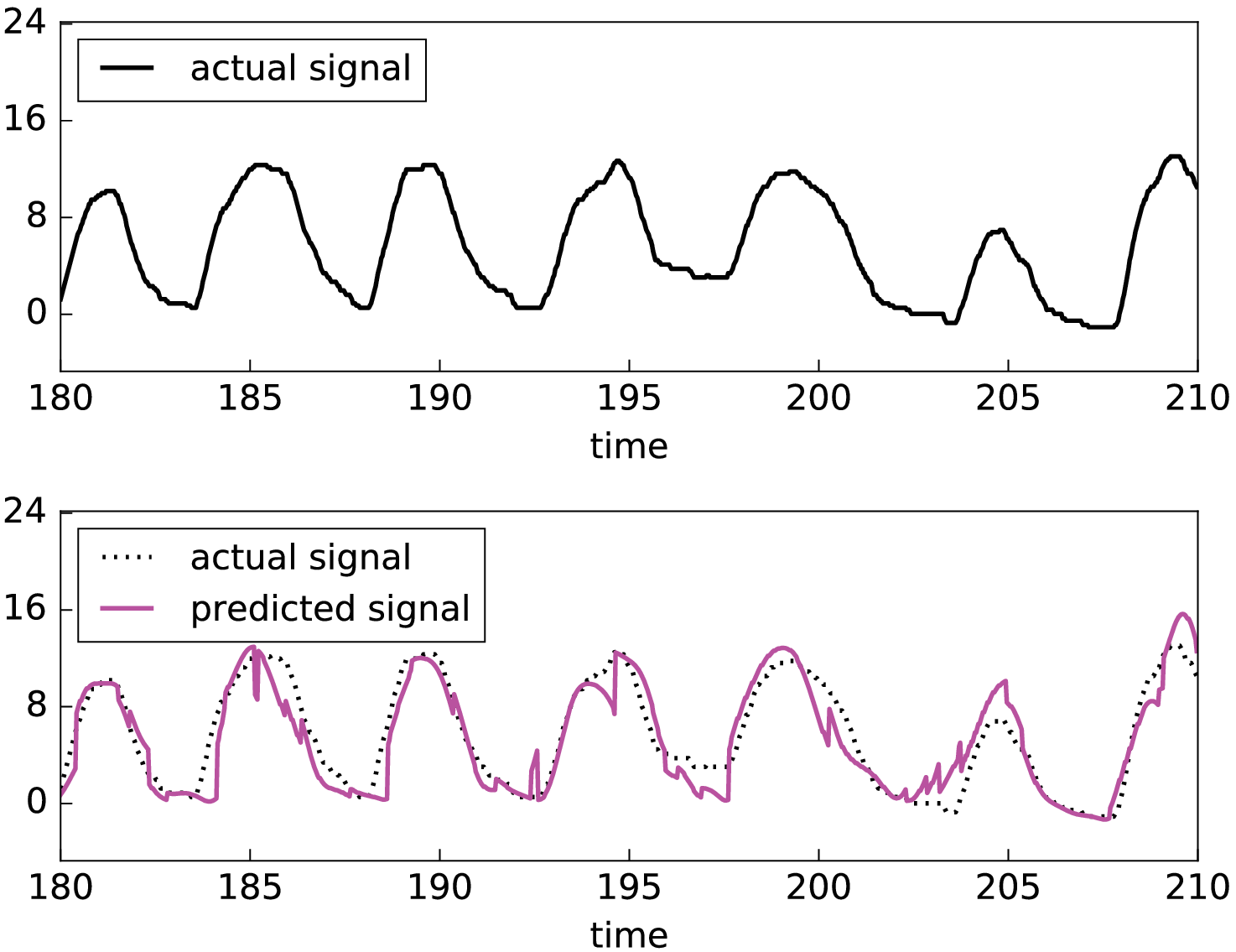}
\caption{Representative prediction result. Predicted signal corresponds to $\hat{x}_{6000},\ldots,\hat{x}_{6999}$ of the serial prediction trials (see \cref{sec:prediction}) using the sample numbered 9. The units of signal value and time are \textup{mm} and \textup{s}, respectively.}
\label{fig:predict2}
\end{figure}

The prediction algorithm was tested using respiratory signals of ten volunteers, measured for 300 s (10000 time points) (\cref{fig:raw}).
For each time point of a signal sample, observations during the past 120 s (4000 points) were used to construct a learning set, and a predictor is formed to forecast the next 0.3 s (10 points) given the previous 3 s (100 points).
In detail, let $N=4000$, $n=100$, $m=10$, and  $\{x_0,\ldots,x_{M-1}\}$ denote a signal sample, where $M=10000$.
For each $t=N+n,\ldots,M-m$, the signal tuple $(x_{t-n-N},\ldots,x_{t-n-1})\in\mathbb{R}^N$ was used to construct a learning set $\mathcal{L}_t\subset \mathbb{R}^n\times\mathbb{R}^m$ as in \cref{sec:learning_set}.
Then $\Psi_{\mathcal{L}_t}(\mathbf{x}_{t})\in\mathbb{R}^m$ was calculated (see \cref{sec:forecast}), where $\mathbf{x}_{t}=(x_{t-n},\ldots,x_{t-1})$.
To evaluate the prediction accuracy, the $m$-th coordinate of $\Psi_{\mathcal{L}_t}(\mathbf{x}_t)$, denoted as $\hat{x}_{t+m-1}$, was compared with the corresponding actual observation $x_{t+m-1}$.
In accordance with the previous studies of predicting respiratory motion \cite{lee14}, the root mean square error (RMSE) (mm)
\begin{equation*}
\sqrt{\frac{\sum_{i=N+n+m-1}^{M-1}\left(\hat{x}_i-x_i\right)^2}{M-N-n-m+1}}
\end{equation*}
was calculated as an indicator of prediction error (\cref{fig:rmse}).
The signal samples with RMSE less than 1.5 mm appeared to be well predictable by our approach (\cref{fig:predict}), while some of the others appeared not to (\cref{fig:predict2}).
Hence the former samples numbered 0, 1, 2, 7, and 8 were selected for the next experiment.

\subsection{Prediction-based RGRT}\label{sec:pRGRT}
\begin{figure}[htbp]
\centering
\subfloat[RGRT]{\label{fig:convRGRT2}\includegraphics[width=0.35\textwidth]{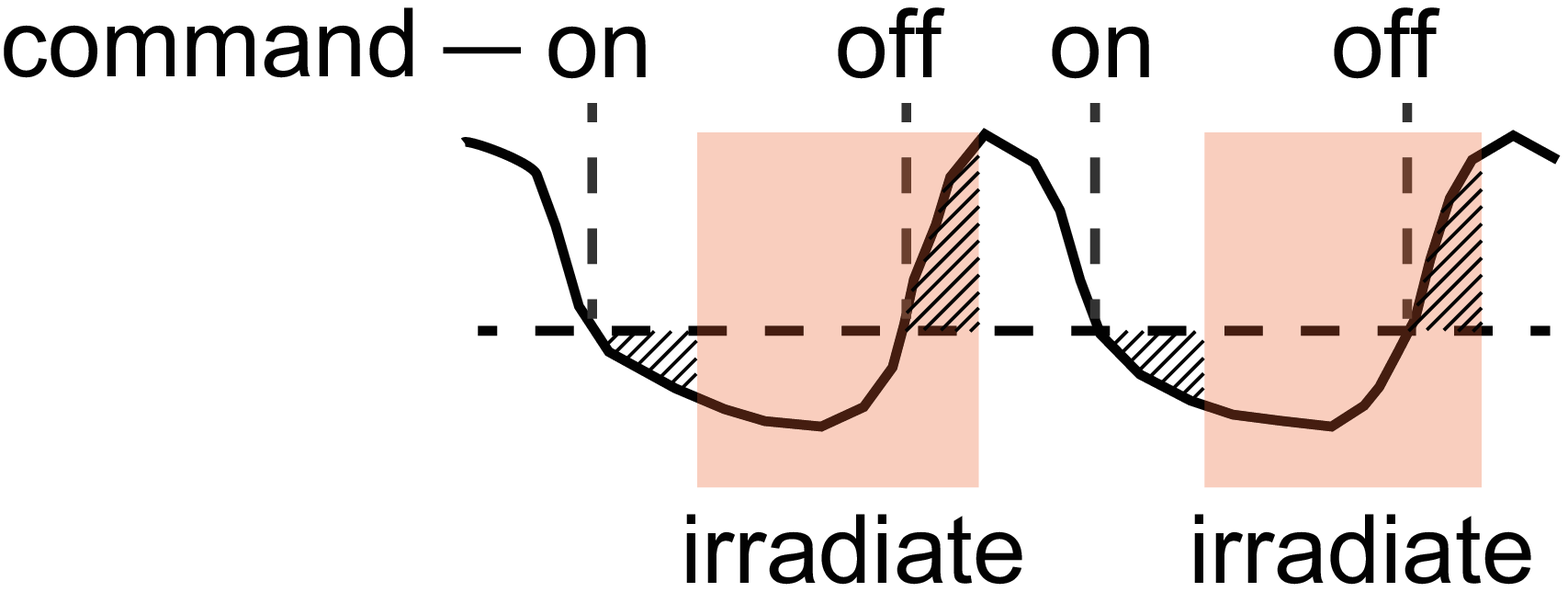}}
\hspace{0.25 in}
\subfloat[pRGRT]{\label{fig:predRGRT}\includegraphics[width=0.406\textwidth]{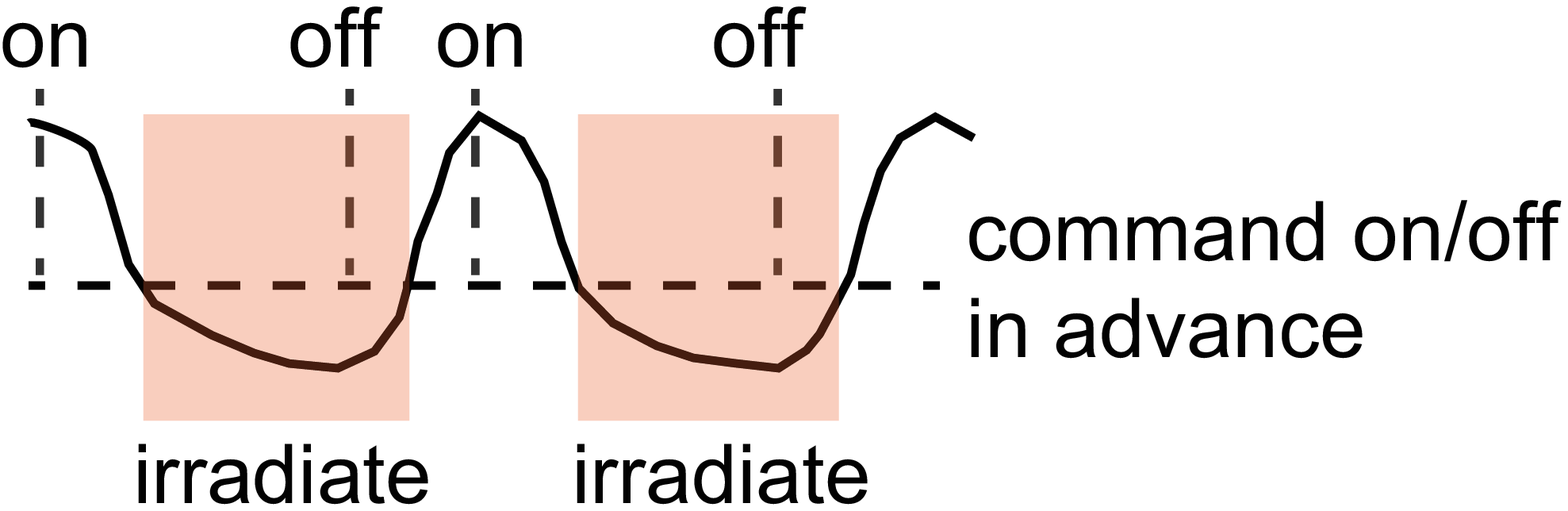}}
\caption{Conventional and prediction-based RGRT systems, denoted here as RGRT and pRGRT, respectively. (a) In RGRT, gate on and off delays cause shifts of gating windows. Stated informally, nErr corresponds to the mean absolute height of the shaded area (\textup{mm}). (b) In pRGRT, gate on and off commands are expected to be sent in advance to compensate for the latencies.}
\label{fig:cpRGRT}
\end{figure}

\begin{figure}[htbp]
\centering
\includegraphics[width=0.50\textwidth]{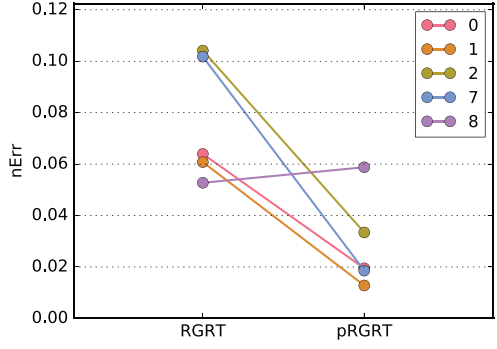}
\caption{nErr values for the selected five samples in the conventional and prediction-based RGRT simulations (denoted here as RGRT and pRGRT, respectively) mimicking the Abches system. The unit of nErr is \textup{mm}.}
\label{fig:nErr_ab}
\end{figure}

\begin{figure}[htbp]
\centering
\includegraphics[width=0.95\textwidth]{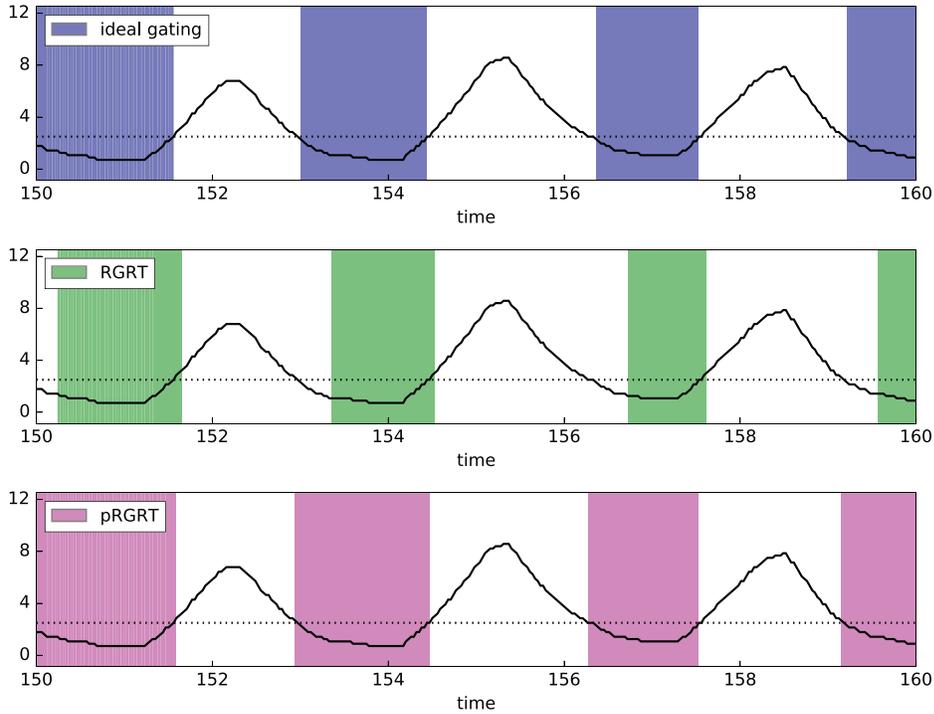}
\caption{Gating windows in the conventional and prediction-based RGRT simulations (denoted here as RGRT and pRGRT, respectively) which mimic the Abches system, using the sample numbered 7. The colored rectangles of RGRT and pRGRT correspond to $\{j: g_j=1\}$ and $\{j: \hat{g}_j=1\}$, respectively, where $5000\le j\le 5333$. The units of signal value and time are \textup{mm} and \textup{s}, respectively.}
\label{fig:7RGRT_ab}
\end{figure}

\begin{figure}[htbp]
\centering
\includegraphics[width=0.95\textwidth]{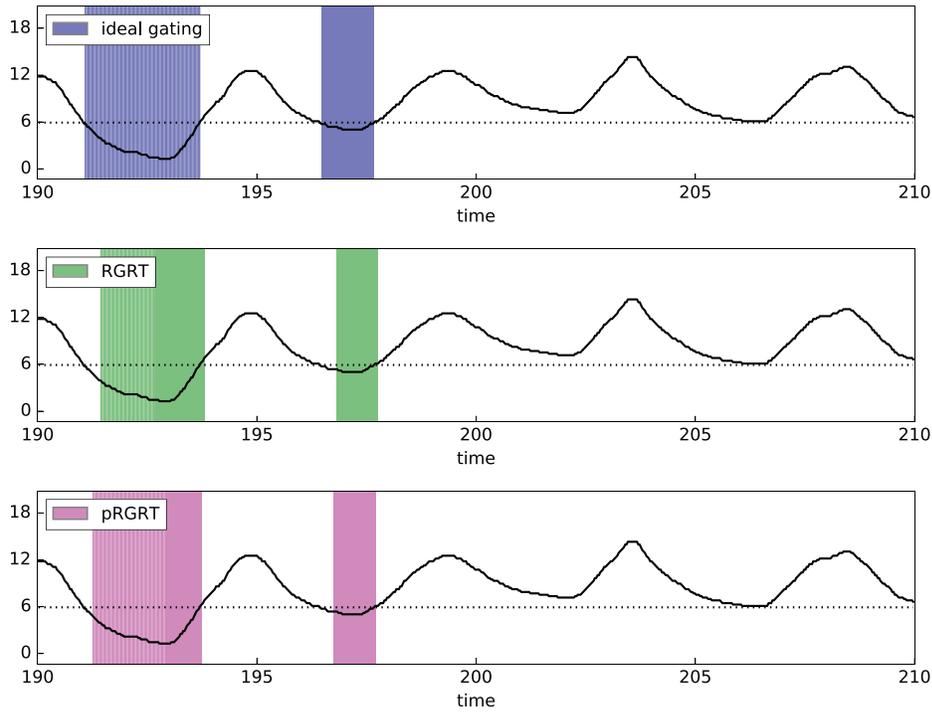}
\caption{Gating windows in the conventional and prediction-based RGRT simulations (denoted here as RGRT and pRGRT, respectively) which mimic the Abches system, using the sample numbered 8. The colored rectangles of RGRT and pRGRT correspond to $\{j: g_j=1\}$ and $\{j: \hat{g}_j=1\}$, respectively, where $6334\le j\le 6999$. The units of signal value and time are \textup{mm} and \textup{s}, respectively.}
\label{fig:8RGRT_ab}
\end{figure}

Our prediction-based gating system, pRGRT, was tested using the selected five signal samples.
In the following experiment, gate on and off delays were set to be 0.336 s and 0.088 s, respectively, in accordance with the Abches system (\cref{tab:delay}).
For each time point $t \ge N+n$ of a sample $\{x_0,\ldots,x_{M-1}\}$, the signal tuple $(x_{t-n-N},\ldots,x_{t-n-1})\in\mathbb{R}^N$ was used to construct learning sets $\mathcal{L}_{1,t}\subset\mathbb{R}^n\times\mathbb{R}^{2m_1+1}$ and $\mathcal{L}_{0,t}\subset\mathbb{R}^n\times\mathbb{R}^{2m_0+1}$ as in \cref{sec:learning_set}, where $M=10000$ (300 s), $N=4000$ (120 s), $n=100$ (3 s), $m_1=12$ (0.336 s), and $m_0=3$ (0.088 s).
We put $\{g_j\}$ and $\{\hat{g}_j\}$ as in \cref{alg:gi} and \cref{alg:gih}, respectively, where $\beta$ was fixed to the median of $\{x_0,\ldots,x_{N-1}\}$.
\begin{algorithm}[H]
\caption{Simulation of conventional RGRT}
\label{alg:gi}
\begin{algorithmic}[1]
\For{$t\gets N+n,\ldots,M-\min\, \{m_0,m_1\}$}
\If{$x_{t-1}<\beta$}
\State $g_{t+m_1-1},\ldots,g_{M-1} \gets 1$,
\Else
\State $g_{t+m_0-1},\ldots,g_{M-1} \gets 0$,
\EndIf
\EndFor
\end{algorithmic}
\end{algorithm}
\begin{algorithm}[H]
\caption{Simulation of pRGRT}
\label{alg:gih}
\begin{algorithmic}[1]
\For{$t\gets N+n,\ldots,M-\min\, \{m_0,m_1\}$}
\If{$G_{\mathcal{L}_{1,t},\mathcal{L}_{0,t}}(t,\beta)=1$}
\State $\hat{g}_{t+m_1-1},\ldots,\hat{g}_{M-1} \gets 1$,
\Else
\State $\hat{g}_{t+m_0-1},\ldots,\hat{g}_{M-1} \gets 0$,
\EndIf
\EndFor
\end{algorithmic}
\end{algorithm}
\noindent
For $j\in S=\{N+n+m_1-1,\ldots,M-1\}$, we assumed that gate on command is executed at $j$
\begin{itemize}
\item  if and only if $g_j = 1$ (in conventional RGRT).
\item  if and only if $\hat{g}_j = 1$ (in pRGRT).
\end{itemize}
In each of the RGRT simulations, let $S_1$ be the set of $j\in S$ at which gate on command is executed, and put $S_0 = S\setminus S_1$.
To quantify possibly inappropriate irradiation during RGRT, the value
\begin{equation*}
\frac{\sum_{j\in S}\left(\chi_{S_1}(j)x_j^++\chi_{S_0}(j)x_j^- \right)}{M-N-n-m_1+1}
\end{equation*}
was calculated and denoted as nErr (normalized error), whose unit is mm.
Here $\chi_S$ represents the characteristic function of a set $S$ defined as
\begin{equation*}
\mathrm{\chi}_S(j) = 
\begin{cases}
\ 1 & \mbox{if}\ \,j\in S\\
\ 0 & \mbox{if}\ \,j\notin S
\end{cases},
\end{equation*}
$x_j^+ = \max\,\{x_j-\beta, 0\}$, and $x_j^- = -\min\,\{x_j-\beta, 0\}$.
Schematic illustrations of nErr and pRGRT are shown in \cref{fig:cpRGRT}.
As a result, nErr values for four out of the five samples decreased in pRGRT (\cref{fig:nErr_ab}).
Regarding the four samples, gating window shifts observed in conventional RGRT appeared to be improved in pRGRT (\cref{fig:7RGRT_ab}).
As for the other sample (numbered 8), considerable baseline drift was observed (\cref{fig:8RGRT_ab}), which is an undesirable feature for gating systems with fixed threshold \cite{eric18}.

\begin{figure}[htbp]
\centering
\includegraphics[width=0.50\textwidth]{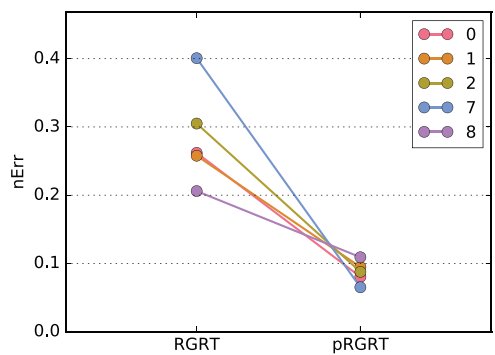}
\caption{nErr values for the selected five samples in the conventional and prediction-based RGRT simulations (denoted here as RGRT and pRGRT, respectively) mimicking the AlignRT system. The unit of nErr is \textup{mm}.}
\label{fig:nErr_Ali}
\end{figure}

\begin{figure}[htbp]
\centering
\includegraphics[width=0.95\textwidth]{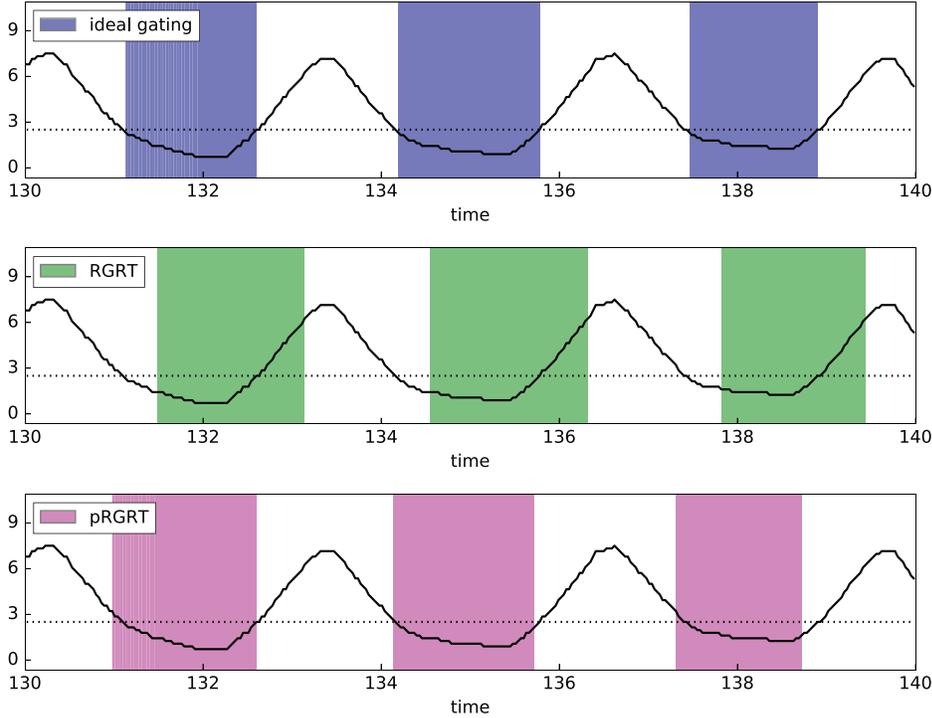}
\caption{Gating windows in the conventional and prediction-based RGRT simulations (denoted here as RGRT and pRGRT, respectively) which mimic the AlignRT system, using the sample numbered 7. The colored rectangles of RGRT and pRGRT correspond to $\{j: g_j=1\}$ and $\{j: \hat{g}_j=1\}$, respectively, where $4334\le j\le 4666$. The units of signal value and time are \textup{mm} and \textup{s}, respectively.}
\label{fig:7RGRT_Ali}
\end{figure}

The above are cases where $m_1\ge m_0$.
To see whether pRGRT works when $m_1< m_0$, similar simulations were performed with gate on and off delays being 0.356 s ($m_1=12$) and 0.529 s ($m_0=18$), respectively, in accordance with the the AlignRT system (\cref{tab:delay}).
The outcome was that nErr values for all the samples decreased in pRGRT (\cref{fig:nErr_Ali}), and gating window shifts in conventional RGRT were ameliorated in pRGRT (\cref{fig:7RGRT_Ali}).

\section{Conclusions}\label{sec:ccl}
In this paper, we proposed a framework to compensate for gate on/off latency during RGRT.
It consisted of two steps: 1) multi-step-ahead prediction and 2) prediction-based gating.
For each step, we devised a specific algorithm to accomplish the task.
Numerical experiments were performed using respiratory signals of a phantom and ten volunteers, and our prediction-based RGRT system, pRGRT, displayed superior performance in not a few of the signal samples.
In some, however, signal prediction and prediction-based gating did not work well, probably because of signal irregularity and/or baseline drift.

The developed method has potential applicability in RGRT, but there are several issues to be addressed, e.g.,
\begin{enumerate}
\item Are there better algorithms for multi-step-ahead prediction?
\item Are there better algorithms for prediction-based gating?
\item Is it possible to deal with baseline drift?
\item Is it possible to provide theoretical foundations to the methods?
\item Is the method valid in a real clinical setting? 
\end{enumerate}
Further studies on these matters would be needed for the system to be of practical use.

\section*{Data availability}
The respiratory signal data used in the current study is available in the Figshare repository (\url{https://doi.org/10.6084/m9.figshare.6290924}).

\section*{Conflicts of Interest}
The authors declare no conflict of interest.

\section*{Funding}
This work was funded by APEX Medical, Inc. (Tokyo, Japan).

\section*{Acknowledgments}
We would like to thank Kazunori Nakamoto (University of Yamanashi) for carefully proofreading a draft of this paper.
We are grateful to Editage (www.editage.jp) for English language editing.

%\newpage
\bibliographystyle{siamplain}
\bibliography{references}
\end{document}

%% file: johno_shared.tex
\usepackage{lipsum}
\usepackage{amsfonts}
\usepackage{graphicx}
\usepackage{epstopdf}
\usepackage[caption=false]{subfig}
\usepackage{algpseudocode}
\ifpdf
  \DeclareGraphicsExtensions{.eps,.pdf,.png,.jpg}
\else
  \DeclareGraphicsExtensions{.eps}
\fi

\newsiamremark{remark}{Remark}
\newsiamremark{hypothesis}{Hypothesis}
\crefname{hypothesis}{Hypothesis}{Hypotheses}
\newsiamthm{claim}{Claim}

\headers{Prediction-based compensation for gate on/off latency}{H. Johno, M. Saito, and H. Onishi}

\title{Prediction-based compensation for gate on/off latency during respiratory-gated radiotherapy\thanks{Accepted by \textit{Computational and Mathematical Methods in Medicine} on October 8, 2018.
}}

\author{Hisashi Johno\thanks{Department of Mathematical Sciences, University of Yamanashi
  (\email{johnoh@yamanashi.ac.jp}).}
\and Masahide Saito\thanks{Department of Radiology, University of Yamanashi 
  (\email{masahides@yamanashi.ac.jp}, \email{honishi@yamanashi.ac.jp}).}
\and Hiroshi Onishi\footnotemark[3]}

\usepackage{amsopn}